\documentclass[floatfix, prl, nopacs, twocolumn,showkeys, preprintnumbers, nofootinbib, superscriptaddress]{revtex4-1}
\usepackage[utf8]{inputenc}
\usepackage[sort&compress]{natbib}
\usepackage{ulem}
\usepackage{overpic}
\usepackage{bm}
\usepackage{times}
\usepackage{amssymb,amsbsy,amsmath,amsfonts}
\usepackage{graphicx}
\usepackage{float}
\usepackage{color}
\usepackage{booktabs}
\usepackage{makecell}

\usepackage{rotating}
\usepackage{srcltx}
\usepackage{slashed}
\usepackage{subfigure}
\usepackage{multirow}
\usepackage{verbatim}
\usepackage{hyperref}
\usepackage{tabularx}
\renewcommand{\arraystretch}{1.2}

\DeclareUnicodeCharacter{3002}{HEREHEREHERE}

\begin{document}

\title{Tribaryons with lattice QCD and one-boson exchange potentials}

\author{Tian-Wei Wu}
\affiliation{School of Science, Shenzhen Campus of Sun Yat-sen University, Shenzhen 518107, China}
\affiliation{School of Fundamental Physics and Mathematical Sciences,Hangzhou Institute for Advanced Study, UCAS, Hangzhou 310024, China}

\author{Si-Qiang Luo}
\affiliation{School of Physical Science and Technology, Lanzhou University, Lanzhou 730000, China}
\affiliation{Lanzhou Center for Theoretical Physics, Lanzhou University, Lanzhou 730000, China}

\author{Ming-Zhu Liu}
\affiliation{School of Space and Environment, Beihang University, Beijing 102206, China}
\affiliation{School of Physics, Beihang University, Beijing 102206, China}

\author{Li-Sheng Geng}\email{lisheng.geng@buaa.edu.cn}
\affiliation{School of Physics, Beihang University, Beijing 102206, China}
\affiliation{Peng Huanwu Collaborative Center for Research and Education, Beihang University, Beijing 100191, China}
\affiliation{Beijing Key Laboratory of Advanced Nuclear Materials and Physics, Beihang University, Beijing 102206, China}
\affiliation{Lanzhou Center for Theoretical Physics, Lanzhou University, Lanzhou 730000, China}
\affiliation{Southern Center for Nuclear-Science Theory (SCNT), Institute of Modern Physics, Chinese Academy of Sciences, Huizhou 516000, China}

\author{Xiang Liu}\email{xiangliu@lzu.edu.cn}
\affiliation{School of Physical Science and Technology, Lanzhou University, Lanzhou 730000, China}
\affiliation{Research Center for Hadron and CSR Physics, Lanzhou University and Institute of Modern Physics of CAS, Lanzhou 730000, China}
\affiliation{Lanzhou Center for Theoretical Physics, Lanzhou University, Lanzhou 730000, China}
\affiliation{MoE Frontiers Science Center for Rare Isotopes, Lanzhou University, Lanzhou 730000, China}

\date{\today}

\begin{abstract}
Motivated by the existence of two-body hadronic molecules composed of $\Omega\Omega$,  $\Omega_{ccc}\Omega_{ccc}$ and $\Omega_{bbb}\Omega_{bbb}$ predicted by lattice QCD simulations, we use the Gaussian expansion method to investigate whether three-body systems composed of $\Omega\Omega\Omega$,  $\Omega_{ccc}\Omega_{ccc}\Omega_{ccc}$ and $\Omega_{bbb}\Omega_{bbb}\Omega_{bbb}$ can bind with the two-body $^1S_0$ interactions provided by lattice QCD. Our results show that  none of the three-body systems bind.
On the other hand, we find that with the one-boson exchange potentials the $\Omega\Omega\Omega$ system develops a bound state,  for which the $^5S_2$ interaction plays an important role. 
Our studies support the existence of the $\frac{3}{2}^+$ $\Omega\Omega\Omega$ bound state and the nonexistence of the $\frac{3}{2}^+$ $\Omega_{ccc}\Omega_{ccc}\Omega_{ccc}$ and $\Omega_{bbb}\Omega_{bbb}\Omega_{bbb}$ bound states, due to the suppressed $^5S_2$ interactions in heavier systems.

\end{abstract}


\maketitle

\noindent{\it Introduction.—}The quark model, as a classification scheme for light-flavor hadrons, was proposed by Gell-Mann~\cite{Gell-Mann:1964ewy} and Zweig~\cite{Zweig:1964jf} in 1964, which was established when the predicted $\Omega$ baryon with the highest strangeness number was observed experimentally~\cite{Barnes:1964pd}. It is often viewed as the first stage in hadron physics. Since 2003, we have witnessed a new stage in hadron physics with the observation of many new hadronic states, such as the charmoniumlike $XYZ$ states and the pentaquark states \cite{Brambilla:2010cs,Liu:2013waa,Chen:2016qju,Chen:2016spr,Oset:2016lyh,Guo:2017jvc,Liu:2019zoy,Brambilla:2019esw,Chen:2022asf}, which have stimulated extensive studies, both theoretically and experimentally. 
Although remarkable progress has been made, a unified understanding of exotic hadronic states is still missing. At present, it is widely acknowledged that one should pay more attention to new configurations, exotic quantum numbers, and special systems in order to better understand the nature of exotic hadronic matter and the nonperturbative strong interaction.

In recent years, fully strange and fully heavy dibaryon systems have attracted  considerable attention.
With increasing computational power, lattice QCD has become the primary force to derive hadron-hadron interactions in a quantitative way from first principles. 
In Ref.~\cite{Gongyo:2017fjb}, the authors  investigated the $\Omega\Omega$ interaction in the $^1S_{0}$ channel, and concluded that there exists a weakly bound state regardless of the Coulomb interaction, which is even shallower than  the deuteron.  
In Ref.~\cite{Lyu:2021qsh}, the existence of  a $^1S_{0}$ $\Omega_{ccc}\Omega_{ccc}$ shallow bound state is predicted while it disappears once the Coulomb interaction is taken into account. Very recently, the existence of a deeply bound $^1S_{0}$ $\Omega_{bbb}\Omega_{bbb}$ state was also predicted~\cite{Mathur:2022nez}. For the $\Omega\Omega$, $\Omega_{ccc}\Omega_{ccc}$, and $\Omega_{bbb}\Omega_{bbb}$ systems,
some of us developed an extended one-boson-exchange (OBE) model to derive their interactions in Ref.~\cite{Liu:2021pdu}, 
and obtained results consistent with those of lattice QCD~\cite{Gongyo:2017fjb,Lyu:2021qsh,Mathur:2022nez}. In Ref.~\cite{Huang:2020bmb}, the authors found the existence of fully heavy dibaryon bound states, $\Omega_{ccc}\Omega_{ccc}$ and $\Omega_{bbb}\Omega_{bbb}$, in the constituent quark model, while the corresponding  fully heavy hexaquark states are found to be above the $\Omega_{ccc}\Omega_{ccc}$ and $\Omega_{bbb}\Omega_{bbb}$ mass thresholds in both the constituent quark models~\cite{Alcaraz-Pelegrina:2022fsi,Lu:2022myk,Weng:2022ohh} and the QCD sum rules~\cite{Wang:2022jvk}.

On the experimental side, studies of fully heavy multiquarks have made important breakthroughs. In 2020, the LHCb Collaboration reported the observation of the first fully heavy tetraquark state, $X(6900)$~\cite{LHCb:2020bwg}. It was later confirmed by  the CMS Collaboration  with a statistical significance of 9.4$\sigma$, and in addition, two new states $X(6600)$ and $X(7200)$ were observed~\cite{X6900}. The ATLAS Collaboration further confirmed the discovery of the LHCb Collaboration~\cite{Evelina}. Clearly, the existence of fully heavy multiquark states can be considered as firmly established.

It is a plausible expectation that tribaryon systems exist, given the predicted existence of dibaryon systems. We note that experimental and theoretical studies of tribaryon  systems other than atomic nuclei and hypernuclei have  continued for many years without conclusive results~\cite{Valcarce:2005em,Park:2018ukx,Garcilazo:1997mf,Garcilazo:1999hh,Mota:1999qp,Valcarce:2001in,Mota:2001ee,Garcilazo:2007ss,Fernandez-Carames:2006uyg, Garcilazo:2022pgt,Garcilazo:2015noa,Garcilazo:2016gkj,Garcilazo:2016ylj, Garcilazo:2020ofz,Garcilazo:2016ams,Garcilazo:2019igo, Zhang:2021vsf,Maezawa:2004va,Sato:2006bw,Sato:2007sb,KEK-PSE549:2009ejj, Yim:2010zza}. 
For example, in Refs.~\cite{Garcilazo:1997mf,Garcilazo:1999hh,Mota:1999qp,Valcarce:2001in,Mota:2001ee}, the authors studied the possible existence of nonstrange tribaryons including $NNN$~\cite{Garcilazo:1999hh,Mota:2001ee}, $NN\Delta$~\cite{Garcilazo:1999hh,Mota:2001ee}, $N\Delta\Delta$~\cite{Mota:1999qp,Mota:2001ee} and $\Delta\Delta\Delta$~\cite{Garcilazo:1997mf,Valcarce:2001in,Mota:2001ee} systems with the relevant two-body potentials.
The strange tribaryons have also been studied, including single strangeness $\Lambda NN$ and $\Sigma NN$~\cite{Garcilazo:2007ss,Fernandez-Carames:2006uyg, Garcilazo:2022pgt}, double strangeness $\Lambda \Lambda N$ and $\Xi NN$~\cite{Garcilazo:2015noa,Garcilazo:2016gkj,Garcilazo:2016ylj, Garcilazo:2020ofz,Garcilazo:2016ams}, and multi strangeness $N\Xi\Xi$~\cite{Garcilazo:2016ams}, $\Omega NN$ and $\Omega\Omega N$~\cite{Garcilazo:2019igo,Zhang:2021vsf}.
An interesting observation is that in the $\Omega NN$ and $\Omega\Omega N$ systems, the $^5S_2$ $\Omega N$ potential  derived from lattice QCD simulations, which can form bound states~\cite{HALQCD:2018qyu}, plays an important role~\cite{Garcilazo:2019igo,Zhang:2020dma,Zhang:2021vsf}.
On the experimental side, a strange tribaryon $S^0(3115)$ was reported in the $^4$He (stopped $K^-$, $p$) reaction, which mainly decays into $\Sigma NN$~\cite{Suzuki:2004ep}. In Ref.~\cite{Maezawa:2004va}, this strange tribaryon is explained as a nonaquark state. Other searches for strange tribaryons have  also been performed~\cite{Sato:2006bw,Sato:2007sb,KEK-PSE549:2009ejj, Yim:2010zza}.

In this paper, motivated by the remarkable progress achieved on studies  of the fully heavy multiquark states from both lattice QCD~\cite{Gongyo:2017fjb,Lyu:2021qsh,Mathur:2022nez} and experiments~\cite{LHCb:2020bwg,X6900,Evelina}, we study the  $\Omega\Omega\Omega$ system, as well as the $\Omega_{ccc}\Omega_{ccc}\Omega_{ccc}$ and $\Omega_{bbb}\Omega_{bbb}\Omega_{bbb}$ systems. Notice that this study differ from the previous works. Regarding the works we mentioned above, they have either different species of baryons ($\Omega NN$, $\Omega\Omega N$, ...) or  different flavors or charges ($pnn$, $\Delta\Delta\Delta$,  $\Xi\Xi\Xi$, ...). To the best of our knowledge, it is the first time that three-body systems that are composed of fully identical flavored baryons have been studied.
The systems we study contain only one species of baryons that  are composed of only one species of quarks and, therefore, have the highest symmetries. This means that we need only the interactions of a pair of identical baryons and the number of allowed configurations is much reduced as well, thus allowing for more robust predictions.

\noindent{\it Most strange tribaryon.—}We adopt the Gaussian expansion method (GEM)~\cite{Kamimura:1988zz,Hiyama:2003cu,Wu:2021ljz} to study the $\Omega\Omega\Omega$ system. To solve the Schr\"{o}dinger equation with GEM, one needs to derive the two-body interactions and  construct the three-body wave functions.     
We note that three-body interactions may play an important role in many-body systems, such as the nucleus. Unfortunately, no empirical information on the three-body interactions is available for the three identical baryons we have studied. Thus, in this exploratory work, we only consider two-body interactions.

\begin{figure}[htpb]
    \centering
\includegraphics[width=8.6cm]{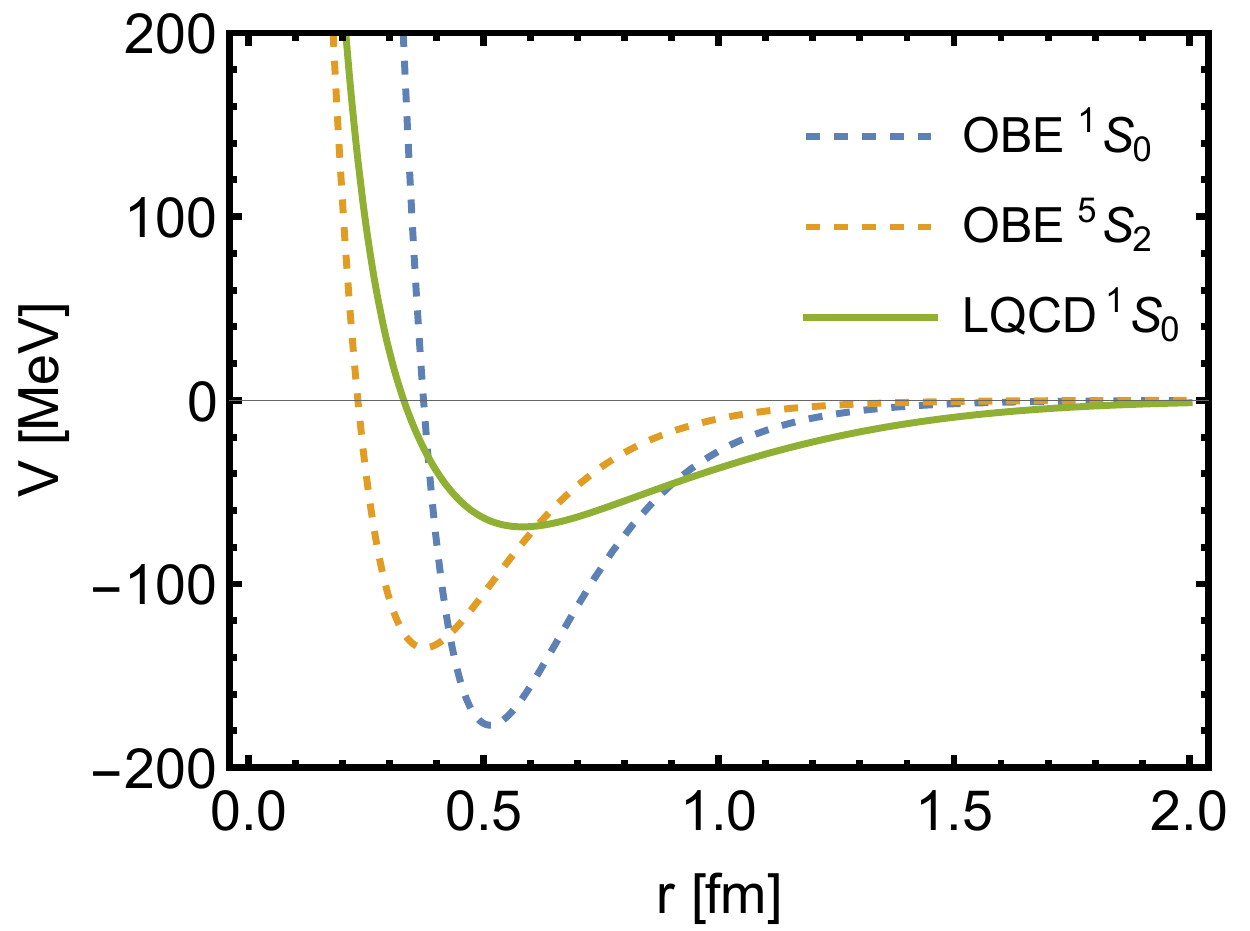}
    \caption{OBE and lattice QCD potentials for the $\Omega\Omega$ system.
    The blue dashed, orange dashed and green solid lines denote the $^1S_0$ OBE, the $^5S_2$ OBE and the $^1S_0$ lattice QCD potentials, respectively.}
    \label{fig:Vomgomg}
\end{figure}

The $\Omega\Omega$ interaction has been derived in lattice QCD~\cite{Gongyo:2017fjb}, where
it was shown that the $S$-wave $\Omega\Omega$ system can bind with a binding energy of $1.6(6)^{+0.7}_{-0.6}$ MeV (without taking into account the Coulomb interaction). In addition to lattice QCD, other methods such as the extended OBE model can also provide the $\Omega\Omega$ interaction~\cite{Liu:2021pdu}. The light meson exchange, including the pseudoscalar($\pi$), scalar($\sigma$) and vector($\rho,\omega$) mesons, can well describe many hadron-hadron interactions,  which is naively extended  to the $\Omega_{ccc}\Omega_{ccc}$ system  by invoking the exchange of the charmonium states  $\eta_c$, $\chi_{c0}$ and $J/\psi$. The   couplings  between $\Omega_{ccc}$ and the  charmonium states are assumed to be proportional to the  couplings between  the nucleon and  the light mesons utilizying the quark model. We emphasize that although  the  OBE model constructed in this way suffers from  relatively large uncertainties, these can be  minimized by fitting to  the lattice QCD binding energies of the $\Omega\Omega$, $\Omega_{ccc}\Omega_{ccc}$, and $\Omega_{bbb}\Omega_{bbb}$ bound states.  In this work, we utilize both interactions to study the $\Omega\Omega\Omega$ three-body system, see Fig.~\ref{fig:Vomgomg} for the potentials.

The $\Omega\Omega$ lattice QCD potentials for the $^1S_0$ channel are expressed with three Gaussian functions $V_L^{^1S_0}(r)=\sum_{i=1}^{3}a_i e^{-b_i r^2}$~\cite{Gongyo:2017fjb}.
Since the $\Omega$ baryon is charged, the Coulomb interaction plays an important role in the $\Omega\Omega$  and $\Omega\Omega\Omega$ systems. The Coulomb potential between a pair of $\Omega\Omega$ is $ V_C(r)=-{\alpha}/{r}$, where $\alpha=1/137$ is the electromagnetic fine structure constant.The lattice QCD simulations provided only the $^1S_0$ potential between the $\Omega\Omega$ pair. As we see later, the $^5S_2$ potential plays an important role in the three-body system as well.

In GEM, a three-body system is studied by solving the three-body Schr\"{o}dinger equation with the three-body wave functions and the Hamiltonian in Jacobi coordinates. For the $\Omega\Omega\Omega$ three-body system, the Schr\"{o}dinger equation is as follows
\begin{equation}
        [T+ V_{\Omega\Omega}(r_1)+ V_{\Omega\Omega}(r_2)+ V_{\Omega\Omega}(r_3)-E]\Psi_J(\vec{r}_c,\vec{R}_c)=0,
\end{equation}
where $c=1-3$ denote the three Jacobi channels, $r_c$($R_c$) are the Jacobi coordinates.
$T$ is the kinetic-energy operator and $V_{\Omega\Omega}$ is the two-body $\Omega\Omega$ interaction. For the details on how to construct the Jacobi coordinates and the three-body kinetic-energy operator, please refer to Ref.~\cite{Hiyama:2003cu}.

The $\Omega\Omega\Omega$ three-body wave function can be written as a sum of three Jacobi channels
\begin{equation}
        \Psi_J(\vec{r}_c,\vec{R}_c)=\sum A_{\alpha}^c \Phi^c_J(\vec{r}_c,\vec{R}_c),
\end{equation}
where $A_{\alpha}^c$ is the expansion coefficients and $\alpha$ is the set of quantum numbers characterizing the wave function in each Jacobi channel.
The wave function of each Jacobi channel reads as
\begin{equation}
\begin{split}
       \Phi^1_J(\vec{r}_1,\vec{R}_1) &=\left[\left[[\chi_3\chi_2]_{s_1}\chi_1\right]_{S}\otimes[\psi_{l_1}(\vec{r}_1)\phi_{L_1}(\vec{R}_1)]_{\Lambda}\right]_{J},\\
        \Phi^2_J(\vec{r}_2,\vec{R}_2)&=\left[\left[[\chi_1\chi_3]_{s_2}\chi_2\right]_{S}\otimes[\psi_{l_2}(\vec{r}_2)\phi_{L_2}(\vec{R}_2)]_{\Lambda}\right]_{J},\\
        \Phi^3_J(\vec{r}_3,\vec{R}_3)&=\left[\left[[\chi_2\chi_1]_{s_3}\chi_3\right]_{S}\otimes[\psi_{l_3}(\vec{r}_3)\phi_{L_3}(\vec{R}_3)]_{\Lambda}\right]_{J},
\end{split}\nonumber
\end{equation}
where $\chi_i$ is the spin wave function of the $i$th particle, $H_{s,S}^c=\left[[\chi_i\chi_j]_{s}\chi_k\right]_{S}$ is the spin wave function of Jacobi channel $c$, $\psi(r_i)\phi(R_i)$ is the spatial wave function, $s$ is the spin of the sub-$\Omega\Omega$ two-body system, $S={3}/{2}$ is the total spin of $\Omega\Omega\Omega$, $l_i$ ($L_i$) is the orbit angular momentum corresponding to $r_i$($R_i$), $\Lambda$ is the total orbit angular momentum built from $l$ and $L$, and $J$ is the total angular momentum built from $\Lambda$ and $S$.

Fermi-Dirac statistics dictates  that only the $^1S_0$ and $^5S_2$ interactions contribute to the formation of an $\Omega\Omega\Omega$ $\frac{3}{2}^+$  state. The spin coupling coefficients of different spin configurations between Jacobi channels $i$ and $j$ for $i\neq j$ are shown in Table \ref{spinfactor}. Note that for $i=j$, the matrix is orthogonal.

\begin{table}[H]
\centering
\caption{Coupling coefficients of different spin configurations between Jacobi channels $i$ and $j$ ($i\neq j$). Here, $H_{s,S}^{c}$ is the spin function,   $s=\{0,2\}$ are alternative spin values of $\Omega\Omega$, and $S={3}/{2}$ is the total spin of $\Omega\Omega\Omega$.}
\renewcommand\arraystretch{1.5}
\begin{tabular*}{43mm}{@{\extracolsep{\fill}}c@{\hskip\tabcolsep\vrule width 0.75pt\hskip\tabcolsep}cc}
\toprule[1.00pt]
\toprule[1.00pt]
 &$H_{0,\frac{3}{2}}^{c=i}$ &  $H_{2,\frac{3}{2}}^{c=i}$ \\
 \hline
 $H_{0,\frac{3}{2}}^{c=j}$& $-\frac{1}{4}$& $-\frac{\sqrt{5}}{4}$ \\
 $H_{2,\frac{3}{2}}^{c=j}$& $-\frac{\sqrt{5}}{4}$ &  $\frac{3}{4}$\\
\bottomrule[1.00pt]
\bottomrule[1.00pt]
\end{tabular*}
\label{spinfactor}
\end{table}

It is important to point out that for the $\Omega\Omega\Omega$ system, the $^5S_2$ potential can play a very important role, even more important than the $^1S_0$ potential. 
This is because the $^5S_2$ partial wave is more strongly coupled to the three-body spin-$3/2$ state than the $^1S_0$ partial wave.
As shown in Table~\ref{spinfactor}, the spin coupling coefficient of different Jacobi channels $i$ and $j$ in the $^5S_2$ partial wave is $\langle H_{2,3/2}^{c=i}|H_{2,3/2}^{c=j}\rangle_{i\neq j}={3}/{4}$ while that in $^1S_0$ is $\langle H_{0,3/2}^{c=i}|H_{0,3/2}^{c=j}\rangle_{i\neq j}=-{1}/{4}$, which means that in the spin space, the coupling between  channels $i$ and $j$ in the $^5S_2$ partial wave is 9 times larger than that in the  $^1S_0$ partial wave.

Once the wave functions are obtained , with either the lattice QCD or OBE $\Omega\Omega$ interactions, one can adopt the GEM~\cite{Hiyama:2003cu} to obtain the binding energies and root-mean-square (rms) radius of the $\Omega\Omega\Omega$ system.

The results for the two-body $\Omega\Omega$ system are summarized in Table~\ref{tab:OOO}, which show that the binding energies and rms radii obtained with the OBE potentials are consistent with those of lattice QCD. With both lattice QCD and OBE potentials, the $\Omega\Omega$ system can bind with a binding energy of $1.4^{+0.9}_{-0.4}$ MeV.
The uncertainties are determined by multiplying a scaling factor to the lattice QCD potential so that the binding energy varies from 1.0 to 2.3 MeV, consistent with the lattice QCD result $1.6^{+0.7}_{-0.6}$ MeV~\cite{Gongyo:2017fjb}.

From the analysis given above, we know that both $^1S_0$ and $^5S_2$ interactions contribute to  the $3/2$ $\Omega\Omega\Omega$ system. Given  that the lattice QCD provided only the $^1S_0$ interaction, we first consider only the $^1S_0$ two-body interaction and find that the $\Omega\Omega\Omega$ system does not bind.
But this result should not be taken too seriously since the $^5S_2$ partial wave plays an important role in the spin configuration ($\langle H_{2,3/2}^{c=i}|H_{2,3/2}^{c=j}\rangle_{i\neq j}=\frac{3}{4}$) and has a significant correlation with the $^1S_0$ partial wave ($\langle H_{0,3/2}^{c=i}|H_{2,3/2}^{c=j}\rangle_{i\neq j}=\frac{\sqrt{5}}{4}$) in the three-body case. 
Actually, with the OBE $^1S_0$ and $^5S_2$ potentials, we find that the three-body $\Omega\Omega\Omega$ system binds  with a binding energy of $5.8^{+2.5}_{-1.2}$ MeV and rms radius $1.9^{+0.1}_{-0.2}$ fm.

Note that the binding energy per baryon of the $\Omega\Omega\Omega$ system is larger than that of the $\Omega\Omega$ system, and consequently its
rms radius is smaller than that of the $\Omega\Omega$ bound state. This is understandable because for the $\Omega\Omega\Omega$ system the $^5S_2$ potential plays an important role, while only the $^1S_0$ potential is relevant for the $\Omega\Omega$ system.

The weights of partial waves and Hamiltonian expectation values of the predicted $\Omega\Omega\Omega$ bound state are given in Table~\ref{tab:HE}, which clearly show that the $^5S_2$ interaction plays a significantly important role in the $\Omega\Omega\Omega$ system. More specifically, the weights of the $^1S_0$ and $^5S_2$ partial waves are about 22\% and 78\%, respectively.

As we mentioned above, since the $\Omega$ baryon is charged, the impact of the Coulomb interaction is worth discussing. We find that the Coulomb interaction in this three-body system affects the binding energy by about 2-3 MeV but does not change the conclusion. Considering the Coulomb interaction, the binding energy and rms radius of the $\Omega\Omega\Omega$ bound state predicted by the OBE model are $2.0$ MeV and $2.3$ fm, respectively.

It is important to discuss where to search for the predicted $\Omega\Omega$ and $\Omega\Omega\Omega$ bound states. In Ref.~\cite{Zhang:2020dma}, the production yield of the $\Omega\Omega$ bound state was estimated using a dynamical coalescence mechanism for the relativistic heavy-ion collisions at $\sqrt{s_{NN}} = 200$ GeV and 2.76 TeV, which turn out to be of the order of $10^{-6}$. In Ref.~\cite{Zhang:2021vsf}, the production yields of $NN\Omega$ and $N\Omega\Omega$ were estimated to be $10^{-7}$ and $10^{-9}$, respectively. Comparing these results, one can estimate the $\Omega\Omega\Omega$ production rate for the relativistic heavy-ion collisions at $\sqrt{s_{NN}} = 200$ GeV and 2.76 TeV, which is of the order of  $10^{-11}$.

\begin{table}[htpb] 
\centering
\caption{Binding energies (BE) and root-mean-square radii ($\langle r\rangle$) of the $\Omega\Omega$ and $\Omega\Omega\Omega$ bound states obtained with lattice QCD (with only $^1S_0$) and OBE  potentials (with both $^1S_0$ and $^5S_2$). BE in MeV and radius $\langle r\rangle$ in fm.}
\renewcommand\arraystretch{1.5}
\begin{tabular*}{86mm}{@{\extracolsep{\fill}}c@{\hskip\tabcolsep\vrule width 0.75pt\hskip\tabcolsep}cccc}
\toprule[1.00pt]
\toprule[1.00pt]
&$\Omega\Omega$(BE)&$\Omega\Omega$($\langle r\rangle$)&$\Omega\Omega\Omega$(BE)&$\Omega\Omega\Omega$($\langle r\rangle$)\\
\midrule[0.75pt]
{LQCD}&$1.41^{+0.89}_{-0.41}$&$3.45^{+0.52}_{-0.62}$&$\cdots$&$\cdots$\\
{OBE}&$1.41^{+0.89}_{-0.41}$&$3.33^{+0.51}_{-0.62}$&$5.84^{+2.48}_{-1.22}$&$1.86^{+0.13}_{-0.19}$\\
\bottomrule[1.00pt]
\bottomrule[1.00pt]
\end{tabular*}
\label{tab:OOO}
\end{table}

\begin{table}[htpb] 
\centering
\caption{Weights of the partial waves and Hamiltonian expectation values (units in MeV) of the $\frac{3}{2}^+$ $\Omega\Omega\Omega$ bound state.}
\renewcommand\arraystretch{1.5}
\begin{tabular*}{86mm}{@{\extracolsep{\fill}}c@{\hskip\tabcolsep\vrule width 0.75pt\hskip\tabcolsep}ccccc}
\toprule[1.00pt]
\toprule[1.00pt]
&$\langle \Psi^{^1S_0}_{3/2}|\Psi^{^1S_0}_{3/2}\rangle$ &$\langle \Psi^{^5S_2}_{3/2}|\Psi^{^5S_2}_{3/2}\rangle$& $\langle T\rangle$&$\langle V^{^1S_0}\rangle$&$\langle V^{^5S_2}\rangle$\\
\midrule[0.75pt]
%
{OBE}&22\% &78\% &52.72& $-15.71$&$-42.84$\\
\bottomrule[1.00pt]
\bottomrule[1.00pt]
\end{tabular*}
\label{tab:HE}
\end{table}

\begin{figure}[t]
    \centering
\includegraphics[width=86mm]{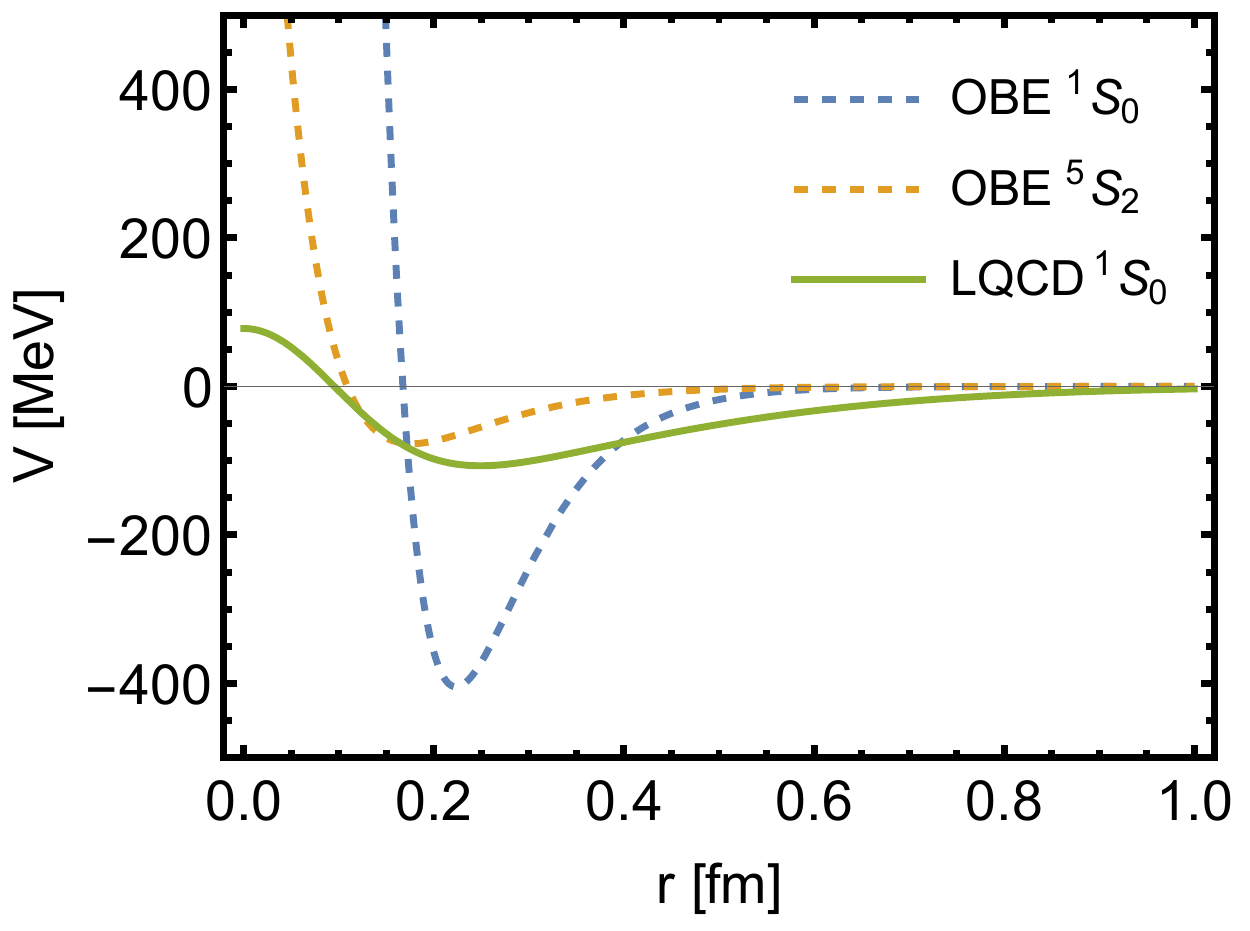}
\includegraphics[width=86mm]{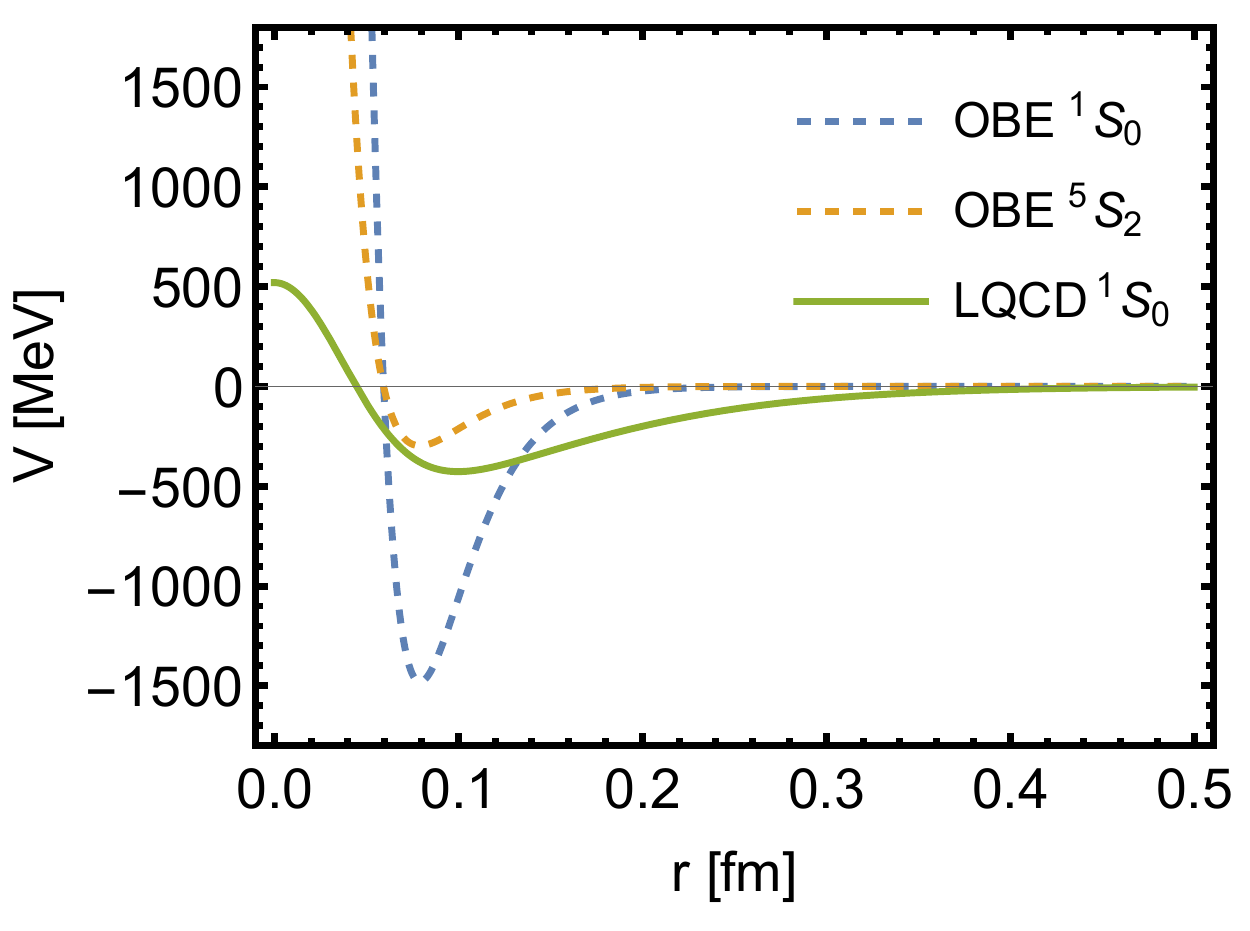}
    \caption{OBE and lattice QCD potentials of the $\Omega_{ccc}\Omega_{ccc}$ (top) and $\Omega_{bbb}\Omega_{bbb}$ (bottom) systems.
    The blue dashed, orange dashed and green solid lines denote the $^1S_0$ OBE, the $^5S_2$ OBE and the $^1S_0$ lattice QCD potentials, respectively.}
    \label{fig:Vomg2}
\end{figure}

\noindent{\it Most charming and beautiful tribaryons.--} It is straightforward to extend the above study to  the $\Omega_{ccc}\Omega_{ccc}$ and $\Omega_{bbb}\Omega_{bbb}$ systems, for which the lattice QCD simulations already provided the $^1S_0$ potentials~\cite{Lyu:2021qsh,Mathur:2022nez} and their OBE counterparts also exist~\cite{Liu:2021pdu}. Note that in Ref.~\cite{Mathur:2022nez} no analytic form of the $\Omega_{bbb}\Omega_{bbb}$ potential was provided. We fitted the lattice QCD potential with a sum of three Gaussian functions as done in Ref.~\cite{Lyu:2021qsh}.   All the lattice QCD potentials and the corresponding OBE potentials are shown in Fig.~\ref{fig:Vomg2}. We note that although the interaction strengths of the lattice QCD potential and those of the OBE potentials are different,  the positions where they become the most attractive are almost the same. The same can be said about the $\Omega\Omega$ potentials shown in Fig.~\ref{fig:Vomgomg}. Such a coincidence indicates that the OBE model must have captured some essential features of the baryon-baryon potentials.

With the above lattice QCD
and the OBE potentials, we can study the two-body and three-body systems composed of $\Omega_{ccc}$ and $\Omega_{bbb}$. As shown in Table~\ref{tab:omegacccomegaccc}, with only the strong interaction the $\Omega_{ccc}\Omega_{ccc}$ bound system can be formed, but it dissolves once the Coulomb interaction is taken into account. On the other hand, the $\Omega_{bbb}\Omega_{bbb}$ system is always bound regardless of the Coulomb interaction. Furthermore, we note that the results obtained with the lattice QCD potentials and those with the OBE potentials are  similar. Nonetheless, none of the $\Omega_{ccc}\Omega_{ccc}\Omega_{ccc}$ and $\Omega_{bbb}\Omega_{bbb}\Omega_{bbb}$ three-body systems can bind, mainly because of the much weaker $^5S_2$ interactions, which are  nontrivial predictions of the present work.

\begin{table}[htpb] 
\centering
\caption{Binding energies (BE) and root-mean-square radii ($\langle r\rangle$) of the $\Omega_{ccc}\Omega_{ccc}$ and $\Omega_{bbb}\Omega_{bbb}$ bound states obtained with OBE and LQCD potentials (BE in MeV and  radius $\langle r\rangle$ in fm.). NC means that the Coulomb interaction is not taken into account, while C means that the Coulomb interaction is considered.}
\renewcommand\arraystretch{1.5}
\begin{tabular*}{86mm}{@{\extracolsep{\fill}}c@{\hskip\tabcolsep\vrule width 0.75pt\hskip\tabcolsep}ccccc}
\toprule[1.00pt]
\toprule[1.00pt]
&&\makecell[c]{$\Omega_{ccc}\Omega_{ccc}$\\(NC)}&\makecell[c]{$\Omega_{ccc}\Omega_{ccc}$\\(C)}&\makecell[c]{$\Omega_{bbb}\Omega_{bbb}$\\(NC)}&\makecell[c]{$\Omega_{bbb}\Omega_{bbb}$\\(C)}\\
\midrule[0.75pt]
\multirow{2}{*}{LQCD}&BE&$5.54$&$\cdots$&$88.7$&79.9\\
&$\langle r\rangle$ &1.14&$\cdots$&0.240&0.245\\
\midrule[0.75pt]
\multirow{2}{*}{OBE}&BE&$5.52$&$\cdots$&$88.6$&$78.4$\\
&$\langle r\rangle$&1.05&$\cdots$&0.198&0.202\\
\bottomrule[1.00pt]
\bottomrule[1.00pt]
\end{tabular*}
\label{tab:omegacccomegaccc}
\end{table}

\noindent{\it Summary.—} Motivated by the existence of $\Omega\Omega$, $\Omega_{ccc}\Omega_{ccc}$, and $\Omega_{bbb}\Omega_{bbb}$ bound states predicted by lattice QCD simulations, we studied the $\frac{3}{2}^+$ $\Omega\Omega\Omega$, $\Omega_{ccc}\Omega_{ccc}\Omega_{ccc}$, and $\Omega_{bbb}\Omega_{bbb}\Omega_{bbb}$ three-body systems with the lattice QCD and OBE potentials. We found that the $\Omega\Omega$, $\Omega_{ccc}\Omega_{ccc}$, and $\Omega_{bbb}\Omega_{bbb}$ systems can also bind with the OBE potentials, with binding energies and rms radii consistent with those of lattice QCD simulations. The repulsive Coulomb interactions plays an important role in these systems especially in the $\Omega_{ccc}\Omega_{ccc}$ system, which is strong enough to break the $\Omega_{ccc}\Omega_{ccc}$ pair bound by the strong force. 

For the three-body systems, we find that the $^5S_2$ partial wave plays a very important role in forming the $\frac{3}{2}^+$ three-body state.
With only the $^1S_0$ lattice QCD potentials, the $\Omega\Omega\Omega$, $\Omega_{ccc}\Omega_{ccc}\Omega_{ccc}$, and $\Omega_{bbb}\Omega_{bbb}\Omega_{bbb}$ three-body systems do not bind.  With the OBE potentials both in  the $^1S_0$ and $^5S_2$ partial waves, the $\Omega\Omega\Omega$ system becomes bound, while the $\Omega_{ccc}\Omega_{ccc}\Omega_{ccc}$ and $\Omega_{bbb}\Omega_{bbb}\Omega_{bbb}$ systems remain unbound
mainly due to the much suppressed attractive $^5S_2$ interaction in the two-body $\Omega_{ccc}\Omega_{ccc}$ and $\Omega_{bbb}\Omega_{bbb}$ systems.
To verify the existence of the $\Omega\Omega\Omega$ bound state, lattice QCD studies of the $^5S_2$ interactions of the $\Omega\Omega$ system will be the key. We hope that the predicted $\Omega\Omega\Omega$ bound state can be searched for in  present and future hadron-hadron colliders.

A particularly interesting discovery of the present work is that even the two-body interactions are attractive and strong enough to form two-body bound states, the three-body systems do not necessarily bind. This is because in three-body systems, spin-spin interactions can play an important role.  The three highly symmetric systems studied in the present work provide an ideal platform to understand the relevance of spin-spin interactions in forming few-body bound states.

\noindent {\it Acknowledgement.—}This work is partly supported by the National Natural Science Foundation of China under Grants No. 11735003, No. 11975041, No. 11961141004,  and the fundamental Research Funds for the Central Universities.  X.L. is supported by the China National Funds for
Distinguished Young Scientists under Grant No. 11825503,
National Key Research and Development Program of China
under Contract No. 2020YFA0406400, the 111 Project under
Grant No. B20063, the National Natural Science Foundation
of China under Grant No. 12247101, the fundamental Research Funds for the Central Universities, and the project for top-notch innovative talents of Gansu province. M.-Z. L. acknowledges support from the National Natural Science Foundation of
China under Grant No. 12105007 and  China Postdoctoral
Science Foundation under Grants No. 2022M710317, and No. 2022T150036.  T.-W. W. acknowledges support from the National Natural Science Foundation of China under Grant No. 12147152 and China Postdoctoral
Science Foundation under Grant No. 2022M723119.

\bibliography{reference}

\end{document}